\documentclass[preprint2]{aastex}

\begin{document}

\title{Breakdown of I-Love-Q universality in rapidly rotating relativistic stars}

\author{Daniela D. Doneva}
\email{daniela.doneva@uni-tuebingen.de}
\affil{Theoretical Astrophysics, Eberhard Karls University of T\"ubingen, T\"ubingen 72076, Germany}
\affil{INRNE - Bulgarian Academy of Sciences, 1784  Sofia, Bulgaria}

\author{Stoytcho S. Yazadjiev}
\affil{Department of Theoretical Physics, Faculty of Physics, Sofia University, Sofia 1164, Bulgaria}
\affil{Theoretical Astrophysics, Eberhard Karls University of T\"ubingen, T\"ubingen 72076, Germany}

\author{Nikolaos Stergioulas}
\affil{Department of Physics, Aristotle University of Thessaloniki, Thessaloniki 54124, Greece}

\author{Kostas D. Kokkotas}
\affil{Theoretical Astrophysics, Eberhard Karls University of T\"ubingen, T\"ubingen 72076, Germany}

\begin{abstract}
It was shown recently, that normalized relations between the
moment of inertia ($I$),
the quadrupole moment ($Q$) and the tidal deformability (Love number) exist
and for slowly-rotating neutron
stars they are almost independent of the equation of state (EOS). We extend  the computation of the $I-Q$ relation to models rotating up to the
mass-shedding limit and show that the \emph{universality of the relations is lost}.
With increasing rotation rate, the  normalized  $I-Q$ relation  departs significantly
from its slow-rotation limit deviating up to $40\%$ for neutron stars and up to $75\%$ for strange stars. The deviation is also EOS dependent and for a broad set of hadronic and strange matter EOS the spread
due to rotation is comparable to the spread due to the EOS, if one considers sequences with fixed rotational frequency. Still, for a restricted
sample of modern realistic EOS one can parameterize the deviations
from universality as a
function of rotation only. The previously proposed $I-$Love$-Q$ relations should thus be used with care, because they lose their universality in astrophysical situations involving compact objects rotating faster than a few hundred Hz.
\end{abstract}


\section{Introduction}

Neutron stars  are at the  intersection of astrophysics,
gravitational physics and nuclear physics  and are thus  excellent
laboratories for investigating strong gravitational fields and the
behavior of matter at very high densities.  Isolated neutron stars or binary
systems  can be a testbed for General Relativity (GR) and other alternative
theories of gravity in the strong field regime, while their oscillations, and binary inspiral and merger phases
are central to  the current efforts
in gravitational wave detection.  Successful detections may allow tighter
constraints  on the equation of state (EOS) of nuclear matter.

The properties of neutron stars, their spacetime curvature  and
gravitational wave emission,  are ultimately connected to their
 internal  structure.  However, at present  there are still large
uncertainties regarding the properties of matter at very high densities.
Some information on the EOS can be indirectly extracted
from the  characteristics of events on the surface of neutron stars
and their exterior spacetime using
astrophysical observations. Steps in this direction are being
taken (see e.g. \citealp{Ozel2013,Steiner2013} and references therein)
and future  observations of X-ray busters and  the properties of accretion disks
in low-mass X-ray binaries may allow the determination of the
radius of neutron stars with  an  accuracy of about $10\%$.
 On the other hand, observations of millisecond pulsars  in binary systems
may allow the
measurement of their moment of inertia with  similar  accuracy
\cite[see][]{Lattimer2005,Kramer2009}. Furthermore, gravitational wave
observations  of the inspiral of binary neutron star systems
will allow for the measurement of their tidal deformability
(see \citealp{Flanagan2008,Hinderer2010,Bini2012,Hotokezaka2013} and \citealp{Radice2013}).

In most cases, such astrophysical measurements are EOS dependent.
This prevents us from accurately determining neutron star properties, such as  mass,
radius and moment of inertia, or distinguishing between neutron stars and strange stars.
The EOS uncertainty does not allow for accurate tests  of alternative theories
of gravity  when the deviations  from General Relativity are of
the same order as the deviations induced by different EOS. The gravitational
wave observations of binary inspirals also face certain problems because of the
degeneracy between the spins and the quadrupole moments of the neutron stars.
A welcome
development in these fields has been the recent discovery of relations between
(normalized expressions of)  the
moment of inertia $I$, the quadrupole moment $Q$ and the  tidal  Love number,
by \citet{Yagi2013,Yagi2013a} (see also \citealp{Urbanec2013},\citealp{Yagi2013c}),
which  are  practically
EOS independent. However, these relations were derived only for non-magnetized stars in the
slow-rotation and small tidal deformation approximations, and the question of their general validity
remained open. The small tidal deformation approximation was later relaxed by \citet{Maselli2013}. They derived similar universal relations for the different phases of neutron star inspirals and concluded that these relations do not deviate significantly from the small tidal deformation approximation. On the other hand, it was shown recently by \citet{Haskell2013}, that the
universality of the $I-$Love$-Q$ relations is broken for strongly magnetized neutron stars with low rotational frequencies.

Here, we investigate the effect of rapid rotation on
the universality of the $I-$Love$-Q$  relations, by computing the
$I-Q$ relation for rapidly rotating relativistic stars (up to the mass-shedding
limit) for a number of different EOS. We show that the EOS universality
breaks down at fast rotation rates, such as those encountered for the fastest
known millisecond pulsars, and the $I-Q$ relation departs significantly from
its slow rotation limit. We should note that other universal relations were derived very
recently by \citet{Pappas2013} which are nearly EOS independent for all rotational
rates of neutron stars within General Relativity. The EOS independence is obtained
when using a normalized angular momentum, instead of the spin frequency.

\section{Formalism}

The metric of a stationary and axisymmetric  spacetime can be written in the following form:
\begin{eqnarray}
ds^2 &=& -e^{2\nu}dt^2 + r^2 \sin^2{\theta} B^2 e^{-2\nu}(d\phi - \omega dt)^2 \nonumber \\
&&+ e^{2(\zeta-\nu)}(dr^2 + r^2 d\theta^2) \label{metric}
.
\end{eqnarray}
The multipole moments of the spacetime are  encoded in
the asymptotic expansion  (at large $r$)  of the metric functions
$\nu$, $B$, $\zeta$ and $\omega$. To leading orders:
\begin{eqnarray}
\nu &=& -\frac{M}{r} + \left(\frac{B_0 M}{3} + \nu_2 P_2 \right)\frac{1}{r^3} + {\cal O} \left(\frac{1}{r^4}\right), \\ \nonumber \\
B &=& 1 + \frac{B_0}{r^2} + {\cal O} \left(\frac{1}{r^4}\right),\\ \nonumber \\
\omega &=& \frac{2J}{r^3} + {\cal O} \left(\frac{1}{r^4}\right),
\end{eqnarray}
where $M$ is the  gravitational  mass, $J$ is the angular momentum, and $P_2$ is a Legendre polynomial, while
$\nu_2$ and $B_0$ are expansion coefficients. The  moment of inertia  for a  uniformly rotating  star
is $I=J/\Omega$, where $\Omega$ is the angular frequency.

The relativistic quadrupole moment is given by:
\begin{eqnarray}
Q = -\nu_2 - \frac{4}{3}\left(\frac{1}{4}+b\right)M^3,
\end{eqnarray}
where $b=B_0/M^2$. A detailed derivation of the above formula is given in the book by \citealp{Friedman2013} and also in the papers by \citealp{Ryan1995,Berti2004,Pappas2012,Pappas2012a}. Our numerical computation was checked against previous results by
\citealp{Berti2004,Pappas2012,Pappas2012a}.

We consider  models of uniformly  rotating stars, obtained with
the {\tt rns} code \citep{Stergioulas1995,Nozawa1998}. Two classes of EOS are examined:  hadronic  EOS, describing
neutron stars and  the MIT bag model of self-bound strange quark matter,
describing strange stars. We choose
six representative neutron star EOS, which are all in agreement with the
observational constraint  of a two solar mass static model, and
two  strange star EOS which also nearly or easily satisfy the two solar mass
constraint, respectively. Our representative set of hadronic EOS are:
\citet{WFF} (WFF2); \citet{AkmalPR} (APR); \citet{Goriely2010} (GCP) and
\citet{Hebeler2010} (HLPS), where the \citet{Douchin2001} crust EOS is used.
Furthermore, we include EOS L by \citet{Pandharipande76}, which is one of the
stiffest proposed EOS, and the zero-temperature limit of the
EOS by \citet{Shen1998a,Shen1998b}.
The two strange star EOS are taken from
\citet{Gondek-Rosinska2008} (denoted by SQSB40
and SQSB60 therein).

\section{Neutron stars}

\begin{figure}[h]
\centering
\includegraphics[width=0.45\textwidth]{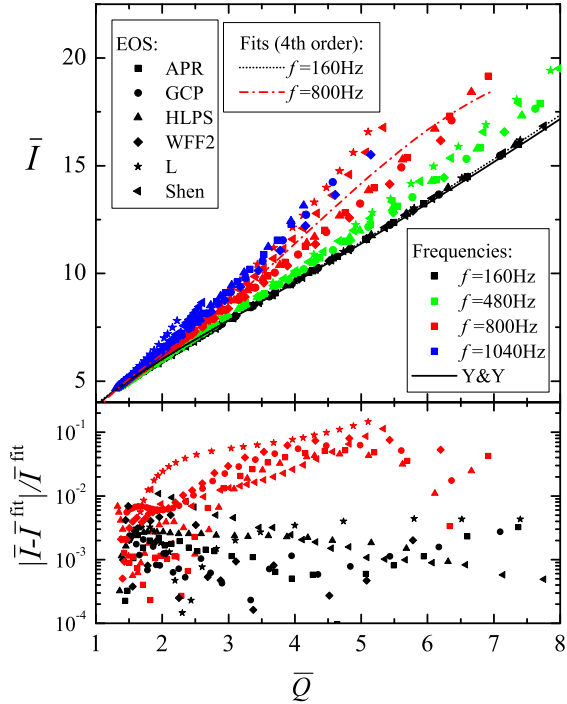}
\caption{{\it Upper panel} The normalized moment of inertia as a function of the
normalized quadrupole moment. Different colors correspond to different
rotational frequencies, while different symbol shapes are used for the
different EOS. The  EOS-independent relation found
by \citet{Yagi2013,Yagi2013a} in the slow rotation limit
is shown as a black line. {\it Lower panel} The deviations from
4th order fits to the data for two representative frequencies
$f=160 {\rm Hz}$ and $f=800 {\rm Hz}$.}
\label{Fig:I(Q)_YY}
\end{figure}

In accordance with \citealp{Yagi2013,Yagi2013a} and \citealp{Haskell2013} we
plot the normalized moment of inertia  ${\bar I}=I/M^3$ as a function of the
normalized quadrupole moment ${\bar Q} = -Q/(M^3 \chi^2)$ where $\chi=J/M^2$.
Notice that we are using the gravitational mass $M$, which is the natural
choice for fast rotating stars. Using the mass
$M_\star$ of a corresponding nonrotating model, as was done by
\citet{Yagi2013,Yagi2013a}, might also be relevant for some astrophysical applications,
because several empirical relations are available that connect the properties of
fast rotating neutron stars (such as the spin frequency at the mass-shedding limit or the
moment of inertia) to the masses and radii of a nonrotating star \cite[see][]{Friedman2013}.
But, we verified that the latter choice leads to
considerably larger deviations from universality for different EOS in the rapidly
rotating case. Fig.
\ref{Fig:I(Q)_YY}  summarizes our results for all
hadronic EOS and for different rotational frequencies $f$.
Different colors correspond to
different fixed rotational frequencies, while the
 universal numerical fit (valid at slow rotation) derived by
\citet{Yagi2013,Yagi2013a} is shown as a solid black line. The deviations
of the numerical results from a fourth-order fit is shown
in the lower panel for two representative rotational frequencies -- in
the slow rotation limit $f=160{\rm Hz}$ and for $f=800{\rm Hz}$, which
is only somewhat larger than the spin frequency of the
fastest known millisecond pulsar $f_{\rm max}\approx700 {\rm Hz}$ \citep{Hessels2006}. The maximum value of $\chi$ is 0.7 for the whole sample of models shown in Fig.~\ref{Fig:I(Q)_YY}.

Several qualitative conclusions can be  drawn, based on Fig. \ref{Fig:I(Q)_YY}.
Most importantly, the ${\bar I}-{\bar Q}$ relation changes considerably with increasing
rotation rate, even for moderately fast rotating neutron stars, such as the fastest
rotating millisecond pulsar currently known.
The deviation from universality is thus significant and can reach
approximately $40\%$ for the selected hadronic EOS\footnote{We define the deviation as the relative difference between ${\bar I}$ for slowly and rapidly rotating models with a fixed value of ${\bar Q}$.}.  On the other hand, even though a universal
${\bar I}-{\bar Q}$
relation does not exist for rapidly rotating neutron stars, we can
consider sequences of fixed rotational
frequency\footnote{\citet{Yagi2013a} suggested that a universal
relation should still exist at fixed rotational frequencies, estimating
a maximum deviation from the slow-rotation limit of the order of 10\% (while
our current results show a maximum deviation of 40\% for hadronic EOS and 75\% for strange matter EOS.).}. Whereas
the ${\bar I}-{\bar Q}$ relation is practically independent of the EOS
in the slow rotation limit (less than 160 Hz in Fig. \ref{Fig:I(Q)_YY}),
for higher rotational frequencies the relation becomes more and more
EOS dependent, and the EOS universality is completely lost at rapid
rotation. \emph{The spread of the data due to different EOS is comparable
to the spread due to rotation} and the deviation from a fourth order
fit $|I-I^{\rm fit}|/I^{\rm fit}$ can reach $15\%$ for millisecond
pulsars (lower panel of Fig. \ref{Fig:I(Q)_YY}).

There is one additional qualitative feature of our results worth mentioning.
As we can see in Fig.
\ref{Fig:I(Q)_YY}, the  spread of the dependences for different equations of
state decreases for lower values of ${\bar I}$ and ${\bar Q}$. Also, the
differences between the dependences for different rotation rates is smaller
in that case. This is an expected effect, since smaller values of the normalized
quantities ${\bar I}$ and ${\bar Q}$ correspond to larger masses and compactness, which
loosely speaking means that we are approaching the black hole limit,
where the ${\bar I}-{\bar Q}$ relation is independent of the internal structure,
as commented by \citet{Yagi2013,Yagi2013a}.

If one restricts attention to
the set of modern EOS that produce 1.4--2.0$M_\odot$ models with
intermediate values for radii $[10.5{\rm km}-12.5{\rm km}]$ (WFF2, APR, GCP and HLPS), as suggested by the observations (see \citealp{Ozel2013,Lattimer2012,Steiner2010}),
then at each fixed rotational frequency a ${\bar I}-{\bar Q}$ relation
that is roughly EOS-independent still holds (with deviations from it
increasing as the mass-shedding limit is approached).

\begin{figure}[t]
\centering
\includegraphics[width=0.48\textwidth]{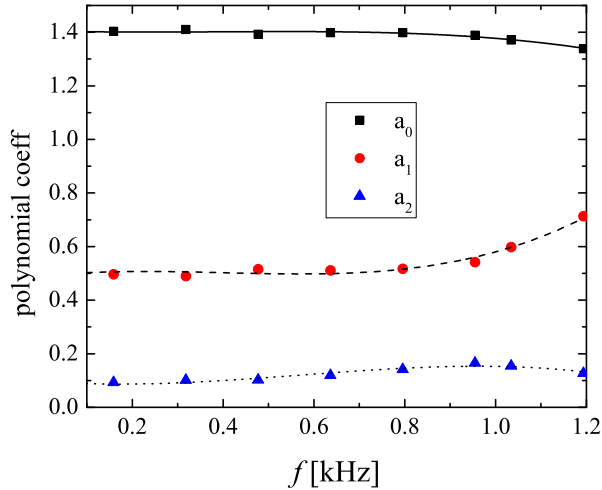}
\caption{The fitting polynomial coefficients $a_i$ as
functions of the rotational frequency. The lines represent a 3rd-order fit to the data.}
\label{Fig:PolynomCoef}
\end{figure}

For the above-mentioned {\it restricted sample} of EOS,
we can quantify how the ${\bar I}-{\bar Q}$ relation changes with
rotation in the following way: we calculated a series of models with
different EOS and central energy densities for several
fixed rotational frequencies. For {\it each} rotational frequency,
we fit the  ${\bar I}-{\bar Q}$ relation with a second-order polynomial of the form
\begin{eqnarray}
\ln {\bar I} = a_0 + a_1 \ln {\bar Q} + a_2 (\ln {\bar Q})^2. \label{Eq:Fit}
\end{eqnarray}
Here, we use a second-order polynomial, in contrast to the fourth-order
fit by \citet{Yagi2013,Yagi2013a} as we find that it is of sufficient
accuracy, while
it is easier to quantify the rotational
dependence of the coefficients $a_i$. A plot of the three fitting coefficients
as a function of the rotational frequency is given in Fig. \ref{Fig:PolynomCoef}.
This dependence can be well approximated with a third-order
polynomial fit of the form\footnote{If we use instead a
4th-order polynomial in Eq. (\ref{Eq:Fit}), as it was done in
\cite{Yagi2013,Yagi2013a},  the error in the fitting of
$a_i(f)$ is considerably larger, especially for low rotational frequencies.}
\begin{equation} \label{Eq:FitCoeff}
a_i = c_{0}+c_{1}\frac{f}{\rm 1kHz}+c_{2}\left(\frac{f}{\rm1kHz}\right)^2
+ c_{3}\left(\frac{f}{\rm1kHz}\right)^3
\end{equation}
In Table 1 we give the results for the coefficients $c_{i}$.

{\renewcommand{\arraystretch}{1.2}
\begin{table} \label{Tbl:Coeff}
\begin{centering}
\begin{tabular}{ccccc}
%\hline
\hline
\hline
\noalign{\smallskip}
$a_i$ & $c_{0}$ & $c_{1}$ & $c_{2}$ & $c_{3}$ \\ \\
\hline
\noalign{\smallskip}
$a_0$  & 1.406  &  -0.051 & 0.154 & -0.131 \\
$a_1$  & 0.489  &  0.183 & -0.562  & 0.471 \\
$a_2$  & 0.098  & -0.136  & 0.463  & -0.273 \\
\noalign{\smallskip}
\hline
\hline
%\hline
\end{tabular}
\end{centering}
\caption{Coefficients in the 3rd-order fit, given by Eq. (\ref{Eq:FitCoeff}), of the $a_i$ coefficients.}
\end{table}
}

\begin{figure}[h]
\centering
\includegraphics[width=0.48\textwidth]{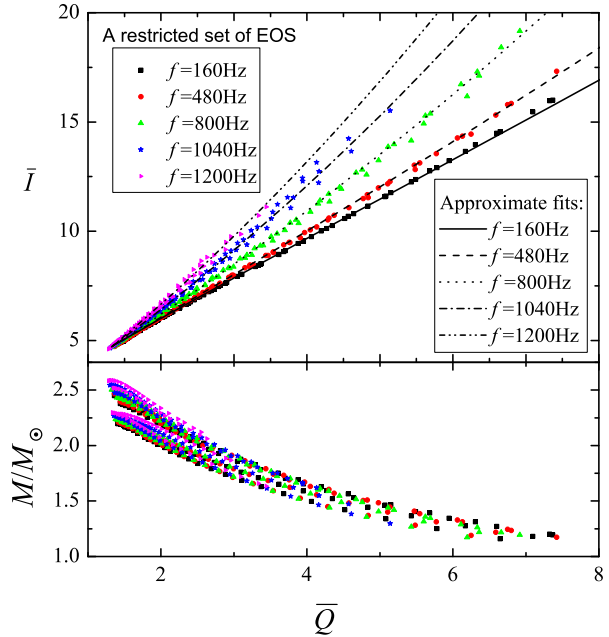}
\caption{ {\it Upper panel}:
Comparison of the numerical data with the polynomial fits
given by Eqs. (\ref{Eq:Fit}) and (\ref{Eq:FitCoeff}) for neutron stars, at several fixed
rotational frequencies. {\it Lower panel}: The mass as a
function of the quadrupole moment for the same models.}
\label{Fig:I(Q)_Fit}
\end{figure}

The question is whether these fitting formulae for the coefficients
$a_i$ are sufficiently accurate.
 Fig. \ref{Fig:I(Q)_Fit} shows that this is indeed the case and
the results in Table 1 can be used
in practice to accurately obtain the ${\bar I}-{\bar Q}$ dependence
(\ref{Eq:Fit}) for our {\it restricted} sample of EOS at different
rotation rates. The deviation from these approximate fits,
defined as $|I-I^{\rm fit}|/I^{\rm fit}$, reaches roughly 2\% for all frequencies below 500Hz and it increases up to 4.5\% for the higher rotational rates. The deviations for the slowly rotating models are mostly due to the error in the approximate fitting formulae (\ref{Eq:Fit}) and (\ref{Eq:FitCoeff}) we derive, while for rotational frequencies above a few hundred Hz it is caused mainly by the spread of the different EOS. In general the least massive models for each frequency, lead to the largest deviations from the approximate fits. One should keep in mind that these deviations depend also on the choice of the set of restricted EOS -- if certain EOS are included or excluded from the set, the percentages might change.

The lower panel in Fig. \ref{Fig:I(Q)_Fit} shows
the  gravitational mass as a function of the
normalized quadrupole moment. For very large masses
(above 2$M_\odot$),   the  ${\bar I}-{\bar Q}$ relation
becomes relatively insensitive to the rotation rate.  However,
at $M=1.4M_\odot$ the relation depends strongly on
rotation.

\subsection{Strange stars} \label{bozomath}

\begin{figure}[t]
\centering
\includegraphics[width=0.48\textwidth]{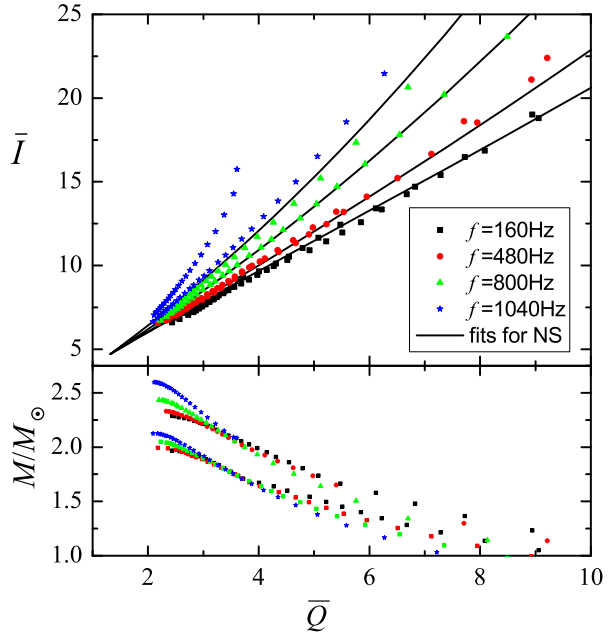}
\caption{{\it Upper panel}: The ${\bar I}-{\bar Q}$ relation for two different strange star EOS
(shown with the same color at a given rotation rate).
For comparison,  the lines show the polynomial fits for neutron stars  at
the same rotational frequencies. {\it Lower panel}: The mass as a
function of the quadrupole moment for the same models.}
\label{Fig:Om_All_QS_Multiple}
\end{figure}

The results for the two strange star EOS we are considering, are shown in
Fig. \ref{Fig:Om_All_QS_Multiple} for several values of $f$.  In order to compare
these with the corresponding relations for neutron stars, we also show
the approximate  polynomial fits for neutron stars derived in the previous section (given by eq. (\ref{Eq:Fit}) and eq. (\ref{Eq:FitCoeff})), for the same values of $f$. The maximum value of $\chi$ for the sample of strange star models considered here, is 1.05.

In the slow rotation limit, the relations for neutron stars and for
both strange star EOS are similar, in accordance with the results in \citet{Yagi2013}
\footnote{The strange matter EOS can be distinguished from the hadronic matter
EOS in the slow rotation limit, if one considers different normalization of the quantities,
as was shown by \citet{Urbanec2013}, where relations between the normalized quadrupole
moment, moment of inertia and compactness were derived.}. However, for rapid rotation the
deviations from the neutron star case become
significant as Fig. \ref{Fig:Om_All_QS_Multiple} shows. The two strange star EOS lead to very different
${\bar I}-{\bar Q}$ relations  (distinct data lines of the same color correspond
to the two difference strange star EOS) and the deviation of the ${\bar I}-{\bar Q}$ relation from its slow-rotation limit reaches approximately $75\%$.  As a result, it is not possible to derive a single fit of the form given by Eq.
(\ref{Eq:Fit}) (at a fixed rotation rate) that would be valid for all strange star EOS.

Another qualitative difference between neutron stars and strange stars
is the following: as we commented above, for neutron star masses near
or above $2M_\odot$ the deviation from universality of
the ${\bar I}-{\bar Q}$ relation decreases significantly,
i.e. the relation becomes relatively insensitive to the rotational
frequency and  to the internal structure. But this is not the case for strange
stars -- even for masses above $2M_\odot$, the ${\bar I}-{\bar Q}$ relations
are quite different for different values of $f$, even though they become more
insensitive to the strange matter EOS. One of the main reasons is that strange
stars reach smaller values of compactness, compared to neutron stars.

\section{Conclusions}

We have calculated the ${\bar I}-{\bar Q}$ relation
for rapidly rotating neutron  stars and strange stars
and showed that it differs significantly from the EOS-independent relation
in the slow-rotation limit found by \citet{Yagi2013,Yagi2013a}.
The deviation from universality can reach about $40\%$ for hadronic EOS and about $75\%$ for strange matter EOS, and for the fastest spinning pulsars currently known, the deviation is around $35\%$ for the two classes of EOS.
The EOS-independence of the ${\bar I}-$Love$-{\bar Q}$ relations is
broken when rapid rotation is considered, because the deviations due
to different EOS are comparable to the deviations induced by rotation if one considers sequences with fixed rotational frequency.
Only for a {\it restricted} set of EOS, that do not include models
with extremely small or large radii, were we still able to find relations
that are roughly EOS-independent at {\it fixed rotational frequencies},
 which can be accurately represented by 2nd-order fits. In addition, we
 find that the complete rotational dependence of the
${\bar I}-{\bar Q}$ relation can be well captured,
if the coefficients in the 2nd-order fits are represented as a 3rd-order
polynomial expansion in the rotational frequency $f$. One should keep in mind that these fits are valid for our chosen restricted set of EOS and if one includes EOS that reach more extreme values of the radii, then the relations would change.

 For strange stars, we find that rapid rotation leads to a completely
different ${\bar I}-{\bar Q}$ relation for different choices of the strange
matter EOS, thus prohibiting more general fits, such as those for neutron
stars.

Our results do not affect the applicability of the $I$-Love-$Q$ relations
in cases where the slow-rotation approximation is sufficient, such as the
inspiral phase of binary neutron star mergers or for the case of some relativistic binaries where
the moment of inertia could be measured. In contrast, our results are important for astrophysical
situations involving compact objects rotating faster than a few hundred Hz.

\acknowledgments
We would like to thank K. Yagi and N. Yunes for critical
comments and suggestions on an early version of the manuscript.
We are grateful to Andreas Bauswein for providing us with an updated collection of recent
EOS tables. DD would like to thank the Alexander von Humboldt Foundation for the support.
KK and SY would like to thank the Research Group Linkage Programme of the Alexander
von Humboldt Foundation for the support. The support from the Bulgarian National
Science Fund under Grant DMU-03/6, by the Sofia University Research Fund under
Grant 33/2013, by the German Science Foundation (DFG) via
SFB/TR7 and the networking support by the COST Action MP1304
is gratefully acknowledged. NS is grateful for the hospitality of the
T\"ubingen group, during an extended stay.

\bibliographystyle{apj}

\end{document}